\documentclass[conference]{IEEEtran}
\IEEEoverridecommandlockouts
\usepackage{cite}
\usepackage{amsmath,amssymb,amsfonts}
\usepackage{algorithmic}
\usepackage{graphicx}
\usepackage{epsfig}
\usepackage{textcomp}
\usepackage{xcolor}

\def\BibTeX{{\rm B\kern-.05em{\sc i\kern-.025em b}\kern-.08em
    T\kern-.1667em\lower.7ex\hbox{E}\kern-.125emX}}
\begin{document}

\title{Ray-Tracing Channel Modeling for LEO Satellite-to-Ground Communication Systems \\}

\author{\IEEEauthorblockN{Jiahao Ning\textsuperscript{1}, Jinhao Deng\textsuperscript{1}, Yuanfang Li\textsuperscript{1}, Chi Zhao\textsuperscript{1}, Jiashu Liu\textsuperscript{1}, \\Songjiang Yang\textsuperscript{2*}, Yinghua Wang\textsuperscript{2}, Jie Huang\textsuperscript{1,2}, Cheng-Xiang Wang\textsuperscript{1,2*}}
\IEEEauthorblockA{\textsuperscript{1}National Mobile Communications Research Laboratory, School of Information Science and Engineering, \\Southeast University, Nanjing 210096, China.\\
\textsuperscript{2}Purple Mountain Laboratories, Nanjing 211111, China.\\
\textsuperscript{*}Corresponding Authors \\
\{Jiahao\_Ning,  jhao\_deng, yuanf\_li, chi\_zhao\_kjt, jshu\_liu\}@163.com, \\\{yangsongjiang, wangyinghua\}@pmlabs.com.cn,
\{j\_huang, chxwang\}@seu.edu.cn}
}

\maketitle

\begin{abstract}
Based on the vision of global coverage for sixth-generation (6G) wireless communication systems, the low earth orbit (LEO) satellite-to-ground channel model for urban scenarios has emerged as highly important for the system design.
In this paper, we propose an LEO satellite-to-ground channel model through shooting and bouncing rays (SBR) algorithm to analyze the channel characteristics. The orbit of LEO is modeled by the simplified general perturbations 4 (SGP4), and an accurate celestial model is applied to calculate the Doppler shift of multipath in a transmission time window of LEO satellite-to-ground communications. Channel characteristics of LEO satellite-to-ground communications such as the root-mean-square (RMS) delay spread, the Doppler shift, and the received power at different times are obtained. The simulation results show that the received power is only significantly noticeable in the transmission time window when the satellite is close to the receiver. Proposed model validates the effectiveness of ray-tracing in actual LEO satellite-to-ground communication scenarios and extends the calculation of the Doppler shift.

\end{abstract}

\begin{IEEEkeywords}
LEO satellite communications, ray-tracing, Doppler shift, coordinate transformation, channel modeling
\end{IEEEkeywords}

\section{Introduction}
	The 6G wireless communication networks present a development vision of global coverage, all spectra, full applications, and strong security, extending from terrestrial mobile communication networks to integrated networks covering the space, air, ground, and sea, achieving global coverage \cite{10054381,10225614,9786750,9237116}. 
	Compared to the synchronous satellite, the LEO satellite can cover remote areas, while offering advantages such as shorter transmission delay, lower path loss, and lower cost.
	To develop LEO satellite communications, the channel model of LEO satellite-to-ground communications is necessary, which can help analyze channel characteristics.

	The channel model method can be divided into stochastic channel models and deterministic channel models. 
	The geometry-based stochastic model (GBSM) combines the geometric structure of the environment with channel characteristics, which can be applied to different scenarios by adjusting the parameters of channel models \cite{10225614,9786750}.
	In \cite{wang2007research}, the authors proposed probability models that fitted the channel characteristics of the LEO satellite-to-ground channels better than the Loo model \cite{1623307}, Corazza model \cite{312773}, and Lutz model \cite{289418}. 
	In \cite{6206541}, the authors proposed a broadband satellite-to-ground channel model simulation platform based on tapped delay line (TDL). However, these channel models are only based on theoretical analysis and have not implemented realistic simulation scenarios.

	The ray-tracing channel modeling method, which is based on geometric optics (GO) and the uniform theory of diffraction (UTD), approximates electromagnetic wave propagation by a ray concept to simulate the reflection, refraction, and diffraction propagation mechanisms in complex environments \cite{SYang2019ISAP}. 
	Due to the requirements of 6G wireless communications for high-accuracy channel models in realistic environments, ray-tracing is an important channel modeling method to achieve the requirements.
	Currently, commercial ray-tracing simulation software, such as Wireless Insite, Ranplan, and Volcano, had certain limitations and their difficult to used in the future 6G LEO satellite-to-ground channel simulations. 
	In  \cite{9474339}, the authors proposed a LEO satellite-to-ground communications based on the ray-tracing channel modeling method, but they did not model the mobility by using actual satellite trajectory. 
	In \cite{10389160}, the authors studied the Doppler shift caused by non-line-of-sight (NLOS) signal on LEO satellite-to-ground communications, but it did not consider the influence of multipath in the Doppler shift. Therefore, developing a LEO satellite-to-ground channel model based on ray-tracing is still a challenging issue.
 
	This paper proposes the transformation method of the satellite coordinate system to the coordinate system of the receiver point in ray-tracing channel modeling, where the actual LEO satellite trajectory is transformed into the LEO satellite-to-ground ray-tracing algorithm in \cite{zhang2024raytracing}. The LEO satellite-to-ground channel characteristics, including path loss, Doppler shift, and delay power spectral density (PSD) are analyzed. The main research contributions and novelties are summarized as follows:

\begin{itemize}
  \item  Based on the two-line element (TLE) data file of the LEO satellite, we apply the SGP4 algorithm to calculate the precise position and velocity of the satellite in the true equator mean equinox (TEME) coordinate system. 

  \item The method to transform the position of satellite from the TEME coordinate system to the earth-centered inertial (ECI) coordinate system is proposed. Additionally, we perform rotations and translations to transform the ECI coordinates into local coordinates, which can be used in the ray-tracing channel modeling algorithm.
  
  \item The precise Doppler shift and RMS delay spread of LEO satellite-to-ground communications at different times in the transmission time window are computed based on the realistic satellite positions and trajectory.

\end{itemize}

The remainder of this paper is structured as follows. Section II outlines the modeling approach for LEO satellites, the large-scale and small-scale fading models are employed. Section III analyzes various characteristics of LEO satellite-to-ground channel, obtaining path loss, RMS delay spread, and Doppler shift of mulpath. Finally, conclusions are drawn in Section IV.

\section{LEO Satellite-to-Ground Channel Modeling}

\begin{figure}[t]
      \centering
      \includegraphics[width=\linewidth]{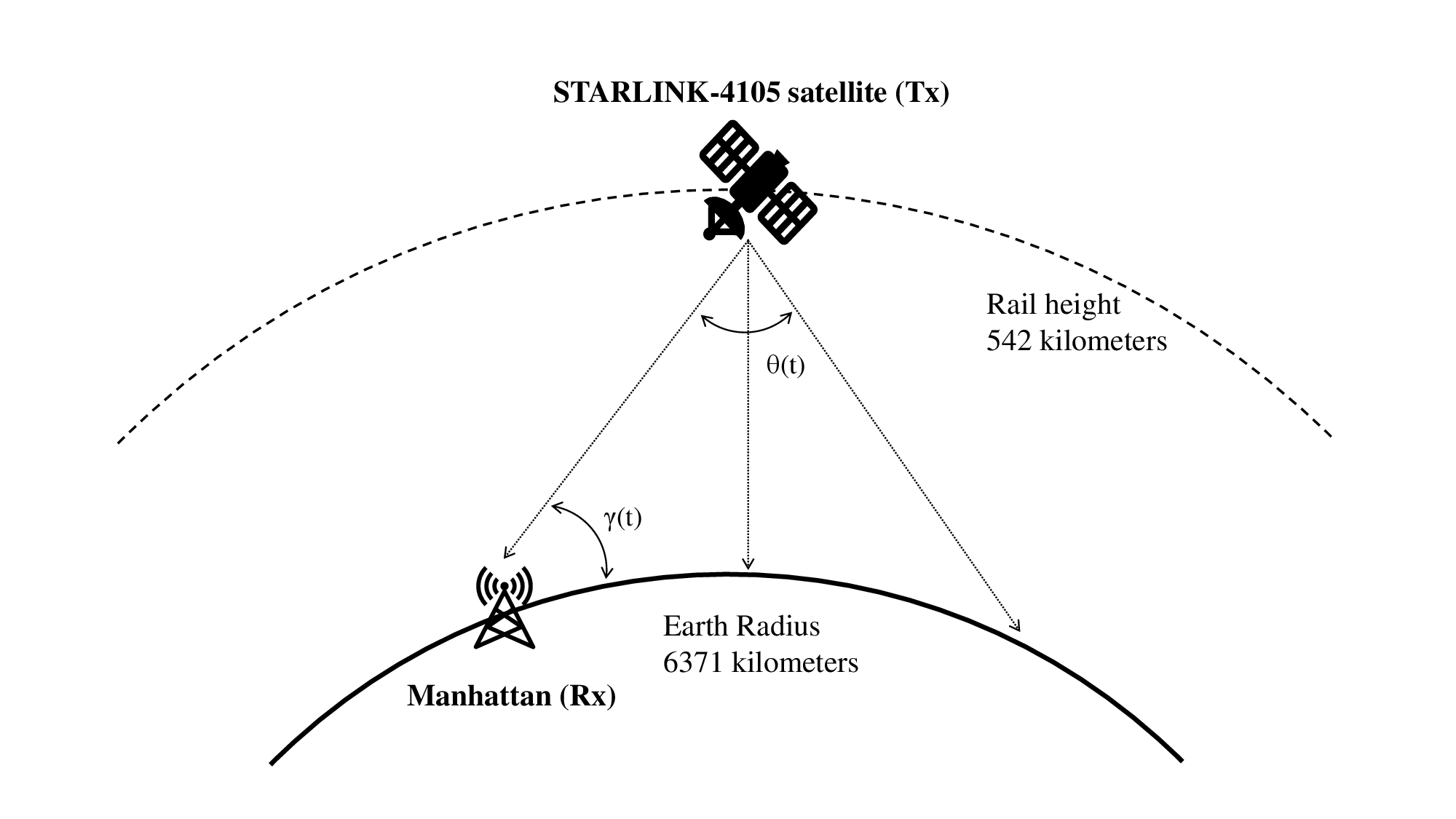}
      \caption{LEO satellite-to-ground communication scenarios.}
      \label{fig1}
\end{figure}

\begin{figure}[t]
      \centering
      \includegraphics[width=\linewidth]{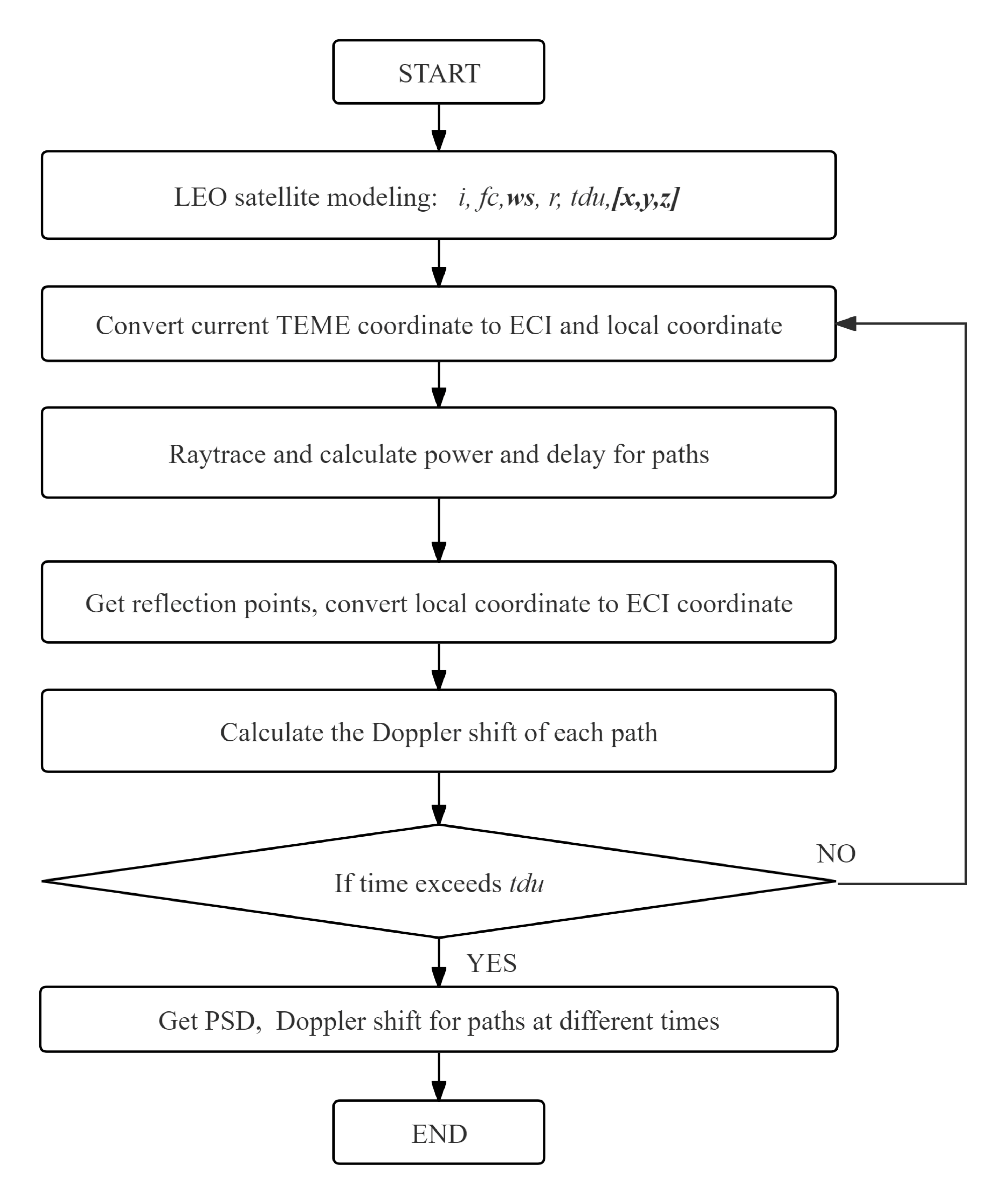}
      \caption{Flow chart of the LEO satellite-to-ground channel modelling method.}
      \label{flow2}
\end{figure}

Fig. \ref{fig1} shows the system model, including the Manhattan city and STARLINK-4105 with its orbit. Fig. \ref{flow2} shows the flow chart of LEO satellite-to-ground channel modeling method.
First, the parameters of LEO satellite should be defined or calculated, which will be explained in the Doppler shift section. Then LEO coordinate systems require a series of transformations to perform ray-tracing, enabling the calculation of channel characteristics at both large and small scales. The transmission time window \textit{$t_{du}$} is calculated in the first step, which shows the effective time period for communication. If time is included in \textit{$t_{du}$}, we change the position of LEO satellite and repeat the above process. If the time exceeds \textit{$t_{du}$}, the ray-tracing process concludes, then we obtain power and Doppler shift of multipath at different times.

\subsection{LEO Satellite-to-Ground Ray-tracing Process}
Given the distance between the LEO satellite and the receiver point in LEO satellite communications, we adopt ray-tracing channel modeling process based on \cite{zhang2024raytracing}. It illustrates that the significant distance between LEO satellite and receiver results in nearly planar propagation of electromagnetic waves in near-ground regions.
When all receiving rays are determined, the power calculation for each ray is initiated. For each ray, the total distance \textit{D} is given by
\begin{equation}
D = D_\mathrm{atmosphere} + D_\mathrm{near-ground}
\end{equation}
where \textit{$D_\mathrm{atmosphere}$} is predetermined as constant when we determine the position of the plane. \textit{$D_\mathrm{near-ground}$} is obtained in the calculation of SBR. We use \textit{D} to compute large-scale losses such as path loss and employ the portion of \textit{$D_\mathrm{near-ground}$} for small-scale fading calculation. 


\subsection{Trajectory of LEO satellite }
Based on TLE data for an LEO satellite, key orbital parameters such as the epoch date, inclination, right ascension of the ascending node, eccentricity, argument of perigee, and mean anomaly, along with other kinematic parameters, are extracted. Utilizing these parameters, the satellite's semi-major axis length and the number of revolutions corresponding to the epoch time are calculated. 

The SGP4 algorithm is utilized for determining the satellite's position and velocity, factoring in disturbances such as the oblateness of the Earth, atmospheric resistance, and gravitational forces exerted by both the Sun and the Moon. The SGP4 algorithm initializes based on the previously extracted key orbital parameters of the LEO satellite. These parameters give a detailed account of the satellite's movement around the Earth. Simultaneously, the algorithm establishes some constants related to geophysical characteristics and the environment, including the average equatorial radius, gravitational constant, third-order gravitational coefficient, atmospheric drag coefficient, and the parameters of the Earth's gravitational field model. These constants embody the influences of Earth’s shape, size, gravitational field distribution, and atmospheric conditions on the satellite’s orbit.

The algorithm solves the Kepler equation to obtain long-period perturbation characteristics and then proceeds with short-period perturbation handling and position-velocity updates. Orthogonal direction vectors relative to the Earth's center are constructed to compute the velocity vector, and through coordinate transformations, the satellite's precise position and velocity in the TEME coordinate system are derived.

Since the dynamic changes in the rotation axis of Earth and its non-uniform shape, an accurate description of the position of satellite and its motion state necessitates coordinate transformations between different reference frames. When converting the coordinates of LEO satellites from the TEME coordinate system to the ECI coordinate system, it is imperative to take into account the Earth's orientation parameters and holistically account for both the long-term and short-term variations in the Earth's rotational axis. According to the IAU 2000 precession model, standard and actual precession angles can be computed, and the transformation from the modern celestial reference frame to the J2000 celestial reference frame can be achieved. Additionally, short-term deviations of the rotation axis from its mean position occur due to nutations induced by the gravitational forces exerted by both the Moon and the Sun. 
According to the IAU 1980 nutation theory, the displacement of the axis of rotation relative to inertial space can be calculated, and convert the celestial coordinates from an instantaneous coordinate system that takes into account short-term variations such as nutation to a coordinate system based on long-term averages.

\begin{figure}[t]
      \centering
      \includegraphics[width=\linewidth]{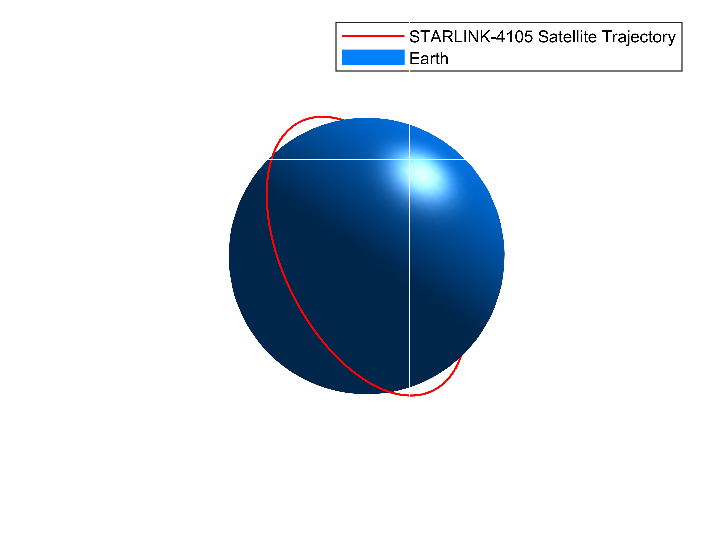}
      \caption{LEO satellite orbital model in ray-tracing.}
      \label{fig2}
\end{figure}

Upon completion of these intricate and precise calculations and transformations, the LEO satellite's orbital coordinates are obtained within the ECI coordinate system. By iteratively executing this computation process, a sequence of ECI coordinates for the satellite at different times is generated, enabling the visualization of the satellite's trajectory evolution in three-dimensional space, as illustrated in Fig. \ref{fig2}.

\subsection{Coordinate transformation of LEO satellite }

\begin{figure}[t]
      \centering
      \includegraphics[width=\linewidth]{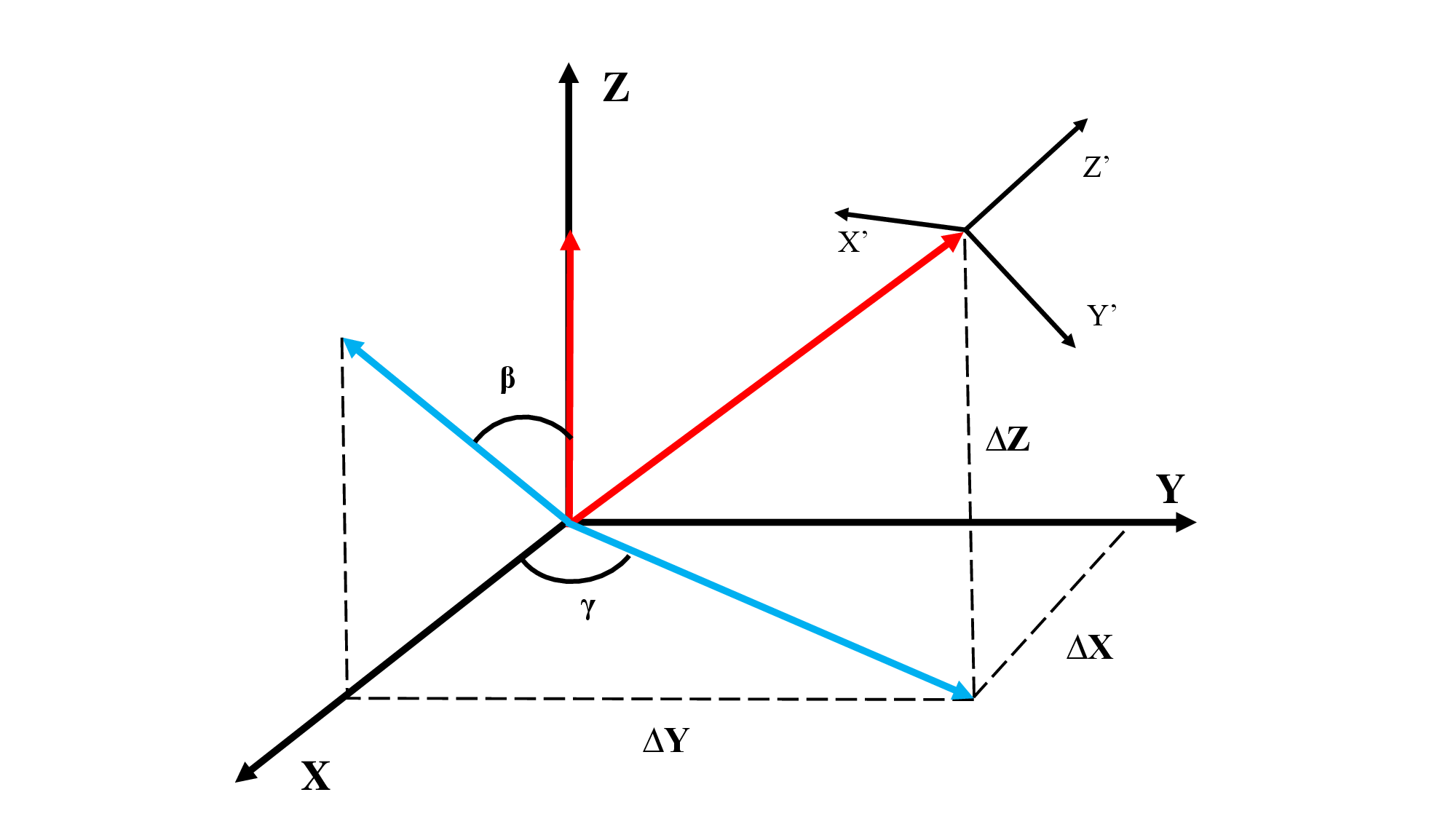}
      \caption{Convert ECI to local coordinate.}
      \label{fig5}
\end{figure}

In ray-tracing channel model, we use the local coordinate system to express the receiver points, while in the previous LEO satellite channel  modeling, the ECI coordinate system (global coordinate system) was used, as seen in Fig. \ref{fig5}. Therefore, before operating the ray-tracing, it is necessary to transform the global coordinates of Manhattan city and the LEO satellite into the local coordinate system.

To convert the global coordinate to the local coordinate, we can consider how to convert the local coordinate to the global coordinate. The main idea is to rotate the vector from Manhattan city to the LEO satellite twice and perform a translation \textit{$\vec{k}$}:
\begin{equation}
\begin{bmatrix}
x \\
y \\
z
\end{bmatrix}
= 
\begin{bmatrix}
\cos \beta & 0 & \sin \beta \\
0 & 1 & 0 \\
-\sin \beta & 0 & \cos \beta
\end{bmatrix}
\begin{bmatrix}
\cos \gamma & -\sin \gamma & 0 \\
\sin \gamma & \cos \gamma & 0 \\
0 & 0 & 1
\end{bmatrix}
\begin{bmatrix}
x' \\
y' \\
z'
\end{bmatrix}
+ \vec{k}
\end{equation}
where $\vec{k}$ is the translation value and angle $\gamma$ represents the angle we rotate the vector from the Earth's center (origin in the global coordinate system) to Manhattan city (denoted as \textit{x', y', z'} in the formula) to the plane xOz. The angle $\beta$ represents the angle we rotate the vector to the z-axis after rotating $\gamma$. 
The above steps transform the local coordinate system into the global coordinate system. Therefore, we can perform the inverse transformation of the above formulas to obtain the local coordinates from the global coordinates.

\subsection{Large-scale and small-scale parameters}
\subsubsection{Path Loss, Rain Attenuation}
The path loss of signal propagation from LEO satellites to the ground is defined by
\begin{equation}
PL=PL_\mathrm{fs}+PL_\mathrm{rain}
\label{eq}
\end{equation}
where \textit{$PL$} is the total path loss, \textit{$PL_\mathrm{fs}$} is the free space path loss and \textit{$PL_\mathrm{rain}$} is the rain attenuation. Based on the Friis Transmission Equation, the \textit{$PL_\mathrm{fs}$} is defined by
\begin{equation}
PL_\mathrm{fs}=32.4+20\log_{10}D(\mathrm{km})+20\log_{10}f_\mathrm{c}(\mathrm{MHz})
\label{eq}
\end{equation}
where \textit{D} is the total distance for each path between LEO satellite and receiver, which is given by (1), and \textit{$f_\mathrm{c}$} is the signal frequency. 
The rain attenuation can be calculated by the rain rate \textit{R} (mm/h),
\begin{equation}
PL_\mathrm{rain}=kR^{\alpha }L
\label{eq}
\end{equation}
where \textit{k} and $\alpha$ are affected by signal frequency. According to different polarization modes, \textit{k} and $\alpha$ have different equation relationships with frequency \textit{$f_{c}$}.  The distance of the signal passing through the rain area is
\begin{equation}
L=[0.00741R^{0.776}+(0.232-0.00018)\sin{\theta_\mathrm{ele}}]^{-1}
\label{eq}
\end{equation}
where $\theta_\mathrm{ele}$ is the elevation angle of the satellite.

\subsubsection{Doppler shift}
\begin{figure}[t]
\centerline{\includegraphics[width=8cm]{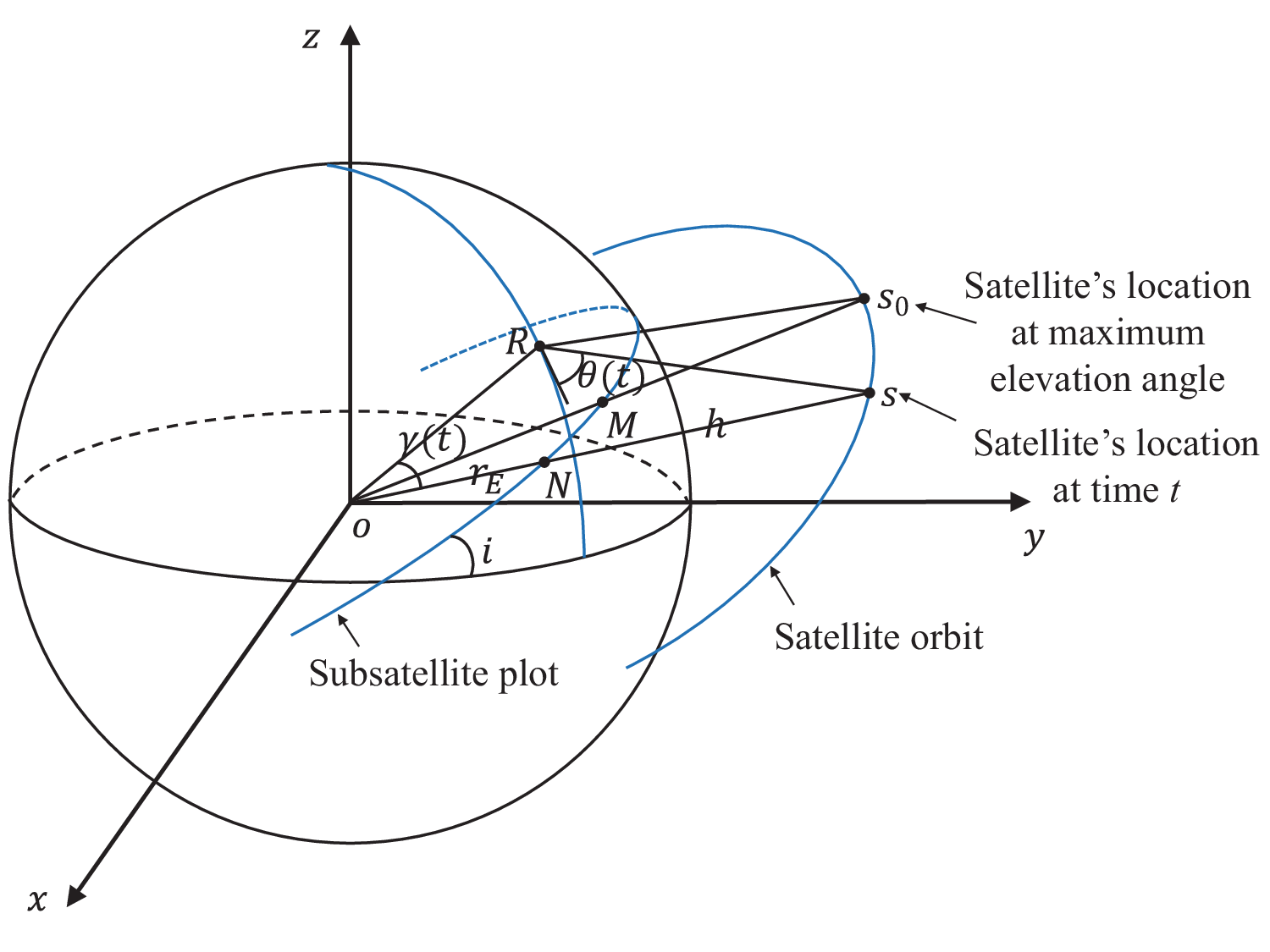}}
\caption{Geometric diagram of LEO satellite and ground terminal.}
\label{fig:geometry}
\end{figure}

Since the velocity of the LEO satellite is much higher than that of the ground terminal, only the Doppler shift caused by LEO satellite movement is considered.
Fig.~\ref{fig:geometry} shows the geometric diagram of LEO satellite and ground terminal, where \textit{R} denotes the ground terminal, \textit{S} denotes the position of satellite at time \textit{t}, \textit{$S_{0}$} denotes the position of satellite when it reaches the maximum elevation angle that ground terminal can observe, \textit{M} denotes the sub-satellite point when satellite reaches the maximum elevation angle that ground terminal can observe, and \textit{N} denotes the sub-satellite point at time \textit{t}. According to the geometry in Fig. \ref{fig:geometry}, the elevation angle can be calculated by
\begin{equation}
\theta (t)=\pi / 2 -\gamma (t) -\angle RSO
\label{eq}
\end{equation}
where $\gamma (t)$ and $\angle RSO$ can be calculated according to the law of cosine.

For a given satellite pass and ground terminal, the maximum elevation angle and the minimum elevation angle are obtained by calculating the elevation angle at each moment.
Based on the geometry in Fig. \ref{fig:geometry}, Doppler shift in the ECI coordinate system can be obtained by the maximum elevation angle, which is proposed by \cite{662636},
\begin{equation}
   f_{D}(t)=-\frac{f_{c}r_\mathrm{E}r\sin (\psi (t)-\psi(t_{0}))\cos \gamma (t_{0}) \cdot \omega _{F}(t)}{c\sqrt{r_{E}^{2}+r^{2}-2r_\mathrm{E}r\cos (\psi (t)-\psi(t_{0}))\cos \gamma (t_{0})} }    
\label{eq}
\end{equation}
where \textit{$r_\mathrm{E}$} is the radius of the earth, \textit{r} is the distance from the satellite to the center of the earth, \textit{$f_\mathrm{c}$} is the carrier frequency, $\psi (t)-\psi(t_{0})$ is the angular distance between \textit{M} and \textit{N}, $\omega _{F}(t)$ is the angular velocity of satellite under ECI frame, and $\gamma (t_{0})$ can be calculated by
\begin{equation}
   \gamma (t_{0})=\cos^{-1} (\frac{r_\mathrm{E}}{r}\cos \theta _\mathrm{max})-\theta _\mathrm{max} 
\label{eq}
\end{equation}
where $\theta_\mathrm{max}$ is the maximum elevation angle.

Based on calculations above, \textit{$t_{du}$} is related to the maximum elevation angle and the minimum elevation angle:

\begin{equation}
   t_{du}\approx \frac{2}{\omega _{s}-\omega _\mathrm{E}\cos i}\cdot \cos^{-1} (\frac{\cos \gamma (t_\mathrm{min})}{\cos \gamma (t_{0}) } )      
\label{eq}
\end{equation}
\noindent where $\omega _{s}$ is the angular velocity of satellite under ECI frame, $\omega _\mathrm{E}$ is the angular speed of the earth's rotation, \textit{i} is the orbital inclination, and \textit{$t_\mathrm{min}$} is the time when ground terminal can just observe the satellite.

\section{Simulation Results and Analysis}
\begin{figure}[thpb]
      \centering
      \includegraphics[width=\linewidth]{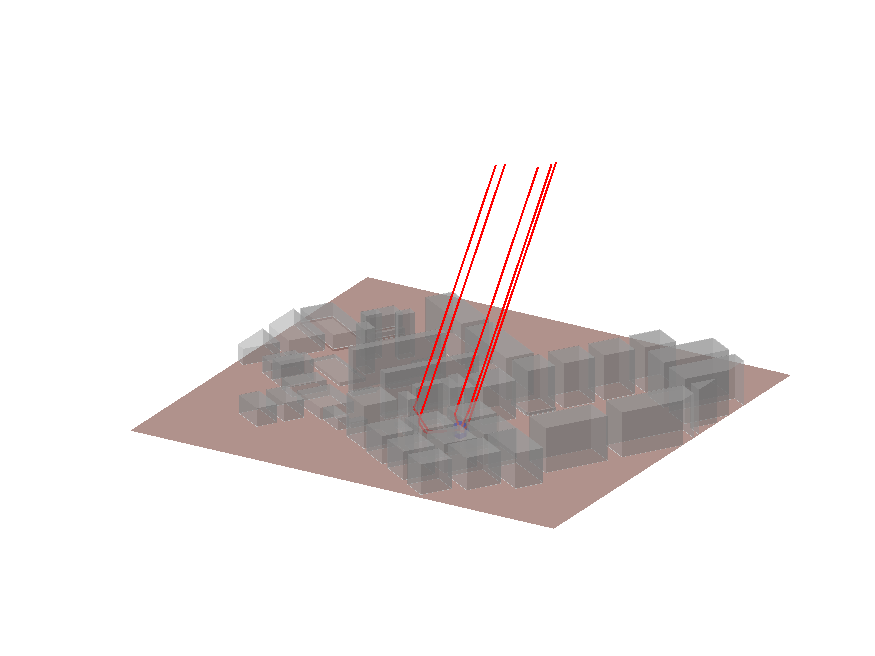}
      \caption{Ray-tracing simulation paths at the 7th minute.}
      \label{paths}
\end{figure}

According to (7), the maximum elevation angle from STARLINK-4105 satellite to Manhattan city is $67.51^{\circ}$, and \textit{$t_{du}$} is 12.76 minutes according to (10).
To observe the multipath tracked at a certain moment, the 7th minute in \textit{$t_{du}$} can be an ideal moment when LEO is closer to Manhattan city in Fig.~\ref{paths}. Simulation parameters can be found in TABLE~I.
\subsection{Path Loss}
\begin{figure}[t]
	\centering
	\includegraphics[width=\linewidth]{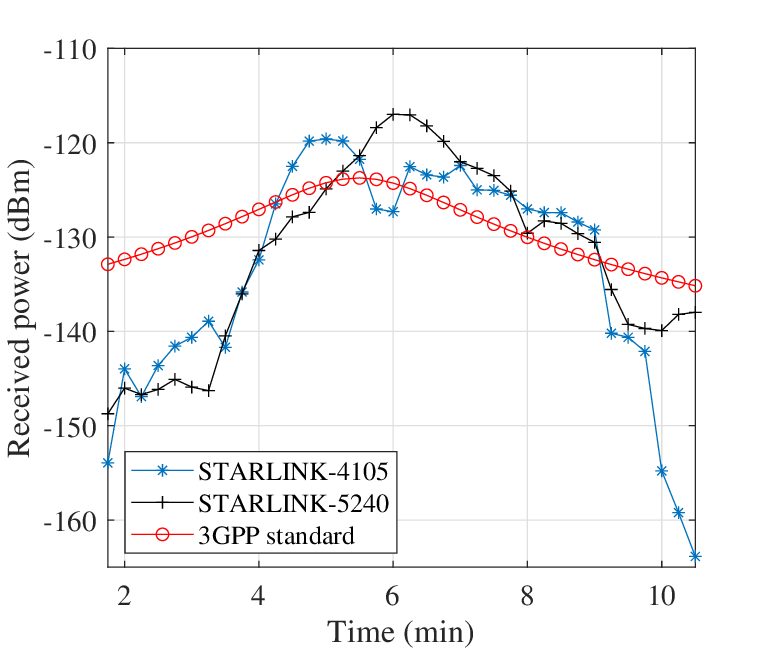}
	\caption{Received power of LEO satellite-to-ground communications.}
	\label{pathloss}
\end{figure}

Fig. \ref{pathloss} shows the variation of the total received power in $t_{du}$ as STARLINK-4105 passes over Manhattan city. Despite fluctuations in the total energy due to building obstructions, the overall trend of increasing and then decreasing satisfies our expectations. 

To verify the accuracy of the results, we compare the result with the 3GPP standard, focusing mainly on path loss. After comparison, we find that the results are quite consistent between the 4th minute and the 9th minute (the satellite elevation angle ranges from $20^{\circ}$ to $67^{\circ}$). However, there is a significant discrepancy when the elevation angle is small. The 3GPP channel model fitted from the measurements considering the diffraction propagation and cannot model blockage accurately, where the received may predicted over the real world. While ray-tracing can provide an accurate depiction of the blockage. After observing the path, we find that when the satellite elevation angle is small, the received rays have undergone two reflections and almost vertically hit the buildings, resulting in small reflection angle and low power.
\begin{table}[t]
\caption{Simulation parameters.}
\centering
\begin{tabular}{|c|c|}
\hline
Parameters & Value  \\
\hline
Carrier frequency (\textit{$f_\mathrm{c}$}) & 2 GHz \\
Transmission power (\textit{$P_\mathrm{t}$}) & 30 dBm \\
Minimum Satellite altitude (\textit{H}) & 542 km \\
Satellite orbit inclination angle (\textit{i}) & $31^{\circ}$ \\
Satellite angular velocity (\textit{$w_\mathrm{s}$}) & 0.066 rad/min  \\
Rain coefﬁcients in 2 GHz (\textit{k/$\alpha$}) & 0.0000847/1.0664 \\
\hline
\end{tabular}
\label{tab:mytable}
\end{table}

Compared with the channel model mentioned in \cite{zhang2024raytracing}, we model the satellite orbit and consider the influence of the orbit on channel characteristics. To verify influence of the orbit, we simulate the path loss of another LEO satellite STARLINK-5240. We find that the path loss under different orbits would have certain deviations, but the overall trend is the same, which proves the robustness of our model.

\subsection{RMS Delay Spread}
\begin{figure}[t]
	\centering
	\includegraphics[width=\linewidth]{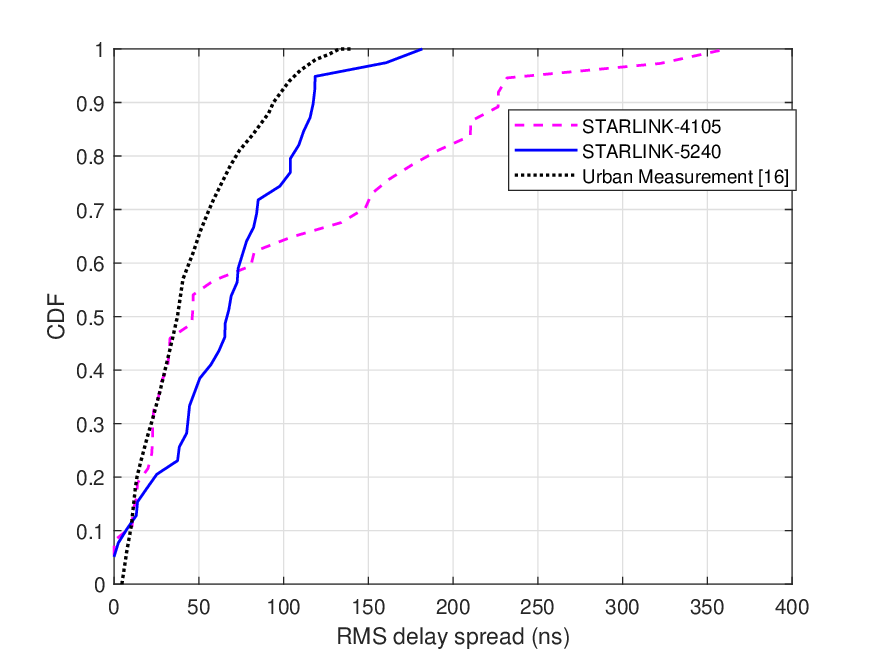}
	\caption{RMS delay spread of the LEO satellite-to-ground in urban scenario compared with measurements \cite{Cid2016TVT}.}
	\label{delay}
\end{figure}

Fig. \ref{delay} shows the LEO satellite-to-ground in urban scenario compared with measurements \cite{Cid2016TVT}. 
The proposed ray-tracing simulation result fits in good agreement with the measurement results.
The multipath effect of the wireless channel can be expressed by the RMS delay spread.  
The trend between measurement and simulation with CDF of RMS delay spread is similar, where the main scatterers, i.e., buildings, impacts on the ray propagation can be captured. 
These comparisons show the accuracy of the proposed method from the channel characteristics perspective.
\subsection{Doppler Frequency Shift}

\begin{figure}[t]
      \centering
      \includegraphics[width=\linewidth]{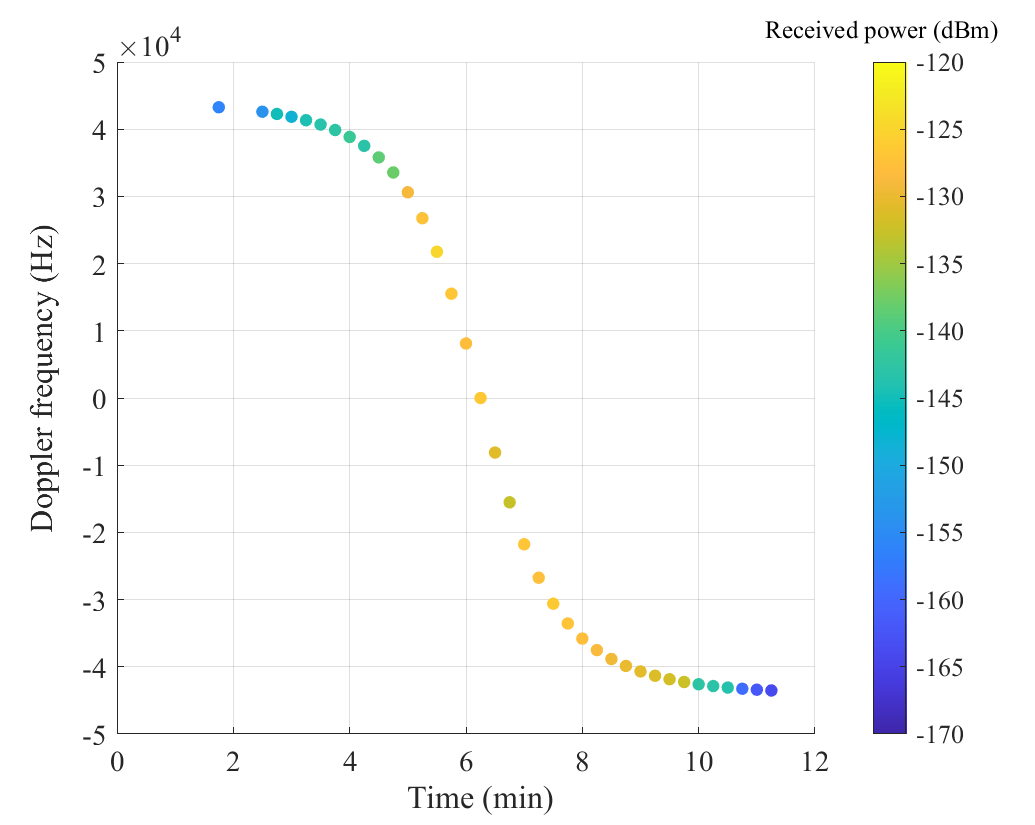}
      \caption{Doppler frequency shift of LEO satellite multipaths.}
      \label{top}
\end{figure}

To more intuitively display the Doppler shift at different time, we observe Fig. \ref{top} by Doppler and time domain. As the Fig. \ref{top} shows, the satellite moves to the position when the ground terminal observes a maximum elevation angle at 6.25 minutes, and the max Doppler shift is 43.7 kHz, which is close to 48 kHz in the 3GPP standard \cite{3gpp2017study}.
The Doppler frequency shift decreases when the satellite moves to the top of the received point.

\section{Conclusions}
In this paper, we have modeled LEO satellites and conducted ray-tracing for the LEO satellite at different times, obtaining the power, delay, Doppler shift, and various large-scale fading of each path at each moment. We have found that the Doppler frequency shift decreases when the satellite moves to the top of the received point. Additionally, due to the influence of buildings and the angle of satellite incidence, the received power has fluctuated to some extent.
Furthermore, we have found that the Doppler shifts of each path are almost equal at the same time (with a deviation of 1-3 Hz) because the distances between reflection points at the same time have been relatively small, resulting in little difference of AoA.

\setcounter{secnumdepth}{0}
\section{Acknowledgment}
This work was supported by the National Natural Science Foundation of China (NSFC) under Grants 61960206006 and 62271147, the Fundamental Research Funds for the Central Universities under Grant 2242022k60006, the Key Technologies R\&D Program of Jiangsu under Grants BE2022067, BE2022067-1, and BE2022067-3, the Start-up Research Fund of Southeast University under Grant RF1028623029, the Fundamental Research Funds for the Central Universities under Grant 2242023K5003, and the Research Fund of National Mobile Communications Research Laboratory, Southeast University, under Grant 2024A05.

\end{document}